\documentclass[twocolumn,showpacs,preprintnumbers,amsmath,amssymb]{revtex4}

\usepackage{graphicx}
\usepackage{dcolumn}
\usepackage{bm}

\begin{document}

\title{Nonlinear Optical Spectroscopy of Photonic Metamaterials}

\author{Evgenia Kim$^{1,2}$}%
\email{evgenia_kim@berkeley.edu}
\author{Feng Wang$^{1,2}$, Wei Wu$^{3}$, Zhaoning Yu$^{3}$, Yuen Ron Shen$^{1,2}$}

\affiliation{%
Department of Physics, University of California, Berkeley, CA, 94720 $^{1}$}%
\affiliation{%
Material Science Division, Lawrence Berkeley National Laboratory, Berkeley, CA, 94720 $^{2}$}%
\affiliation{%
Quantum Science Research, HP Labs, Hewlett-Packard, Palo Alto, CA, 94304 $^{3}$ }%

\begin{abstract}
{We have obtained spectra of second-harmonic generation, third
harmonic generation, and four-wave mixing from a fishnet
metamaterial around its magnetic resonance. The resonant behaviors
are distinctly different from those for ordinary materials. They
result from the fact that the resonance is plasmonic, and its
enhancement appears through  the local field in the nanostructure.}
\end{abstract}

\maketitle

Optical metamaterials with nanoscale metal building blocks have
been studied extensively in recent years
~\cite{pendry}-~\cite{dolling}. With each metal unit much smaller
than optical wavelength, they can be viewed as continuous media
and the metal units act as 'artificial molecules'.  The optical
properties of metamaterials can be engineered by proper design of
the 'artificial molecules' and they can exhibit unusual behavior
nonexistent in nature, such as negative refractive indices. While
linear optical properties of metamaterials have been well
investigated ~\cite{smith}-~\cite{dolling}, nonlinear optical
properties began to attract interest only recently
~\cite{linden}-~\cite{huang}. Such interest stems from possible
strong enhancement of nonlinear response from plasmon resonances
of the metal nanostructures together with spectral tunability
offered by design of 'artificial molecules'. A crucial aspect in
understanding nonlinear optical properties of metamaterials is
their spectral responses, which have not yet been reported.

In this letter, we present the first spectroscopic study of second
harmonic generation (SHG), third harmonic generation (THG) and
four-wave mixing from a metamaterial comprising a monolayer of
"fishnet" structure~\cite{szhang},~\cite{osgood}. It was designed
to have negative refractive index in the near-IR range, with a
magnetic resonance around 1.55 $\mu$m~\cite{ourapa}. The spectra
of SHG and THG with the fundamental input scanned over the
magnetic resonance were obtained with different fundamental and
harmonic polarizations. Resonant enhancement were clearly
observed. Interestingly, the observed resonances are much sharper
than that in linear absorption. This is distinctly different from
typical molecular cases, where resonant excitation at the
fundamental wavelength yields the same resonance spectrum in
linear and nonlinear responses. Such difference originates from
the fact that, unlike molecular resonances, the plasmon resonances
in metal nanostructures are collective oscillations and their
resonance enhancement appears through the local field effect in
the nonlinear processes.

The  measurements were carried out on a "fishnet" metamaterial
composed of two silver sheets with hole arrays separated by a
SiO$_{2}$ layer. It was fabricated using combination of
nanoimprint lithography (NIL) and electron-beam lithography (EBL)
~\cite{wu}. The SEM image of the structure is shown in Fig. 1a and
the structural configuration of the broad wire and its dimensions
in Fig. 1b. Linear optical response of this metamaterial exhibits
a magnetic resonance at ~1.55 $\mu$m when the magnetic field
component of the input wave threads the loop formed by linking the
broad metal wires of the two layers, as indicated in Fig. 1b by
the black arrows ~\cite{lagarkov},~\cite{podolsky}.
Correspondingly, the effective refractive index is negative in the
wavelength range of 1.45 to 1.6 $\mu$m. The measured linear
transmittance and reflectance spectra are presented in Fig. 1c.

\begin{figure}[!h]
\begin{centering}
\includegraphics
[width=8.5cm]{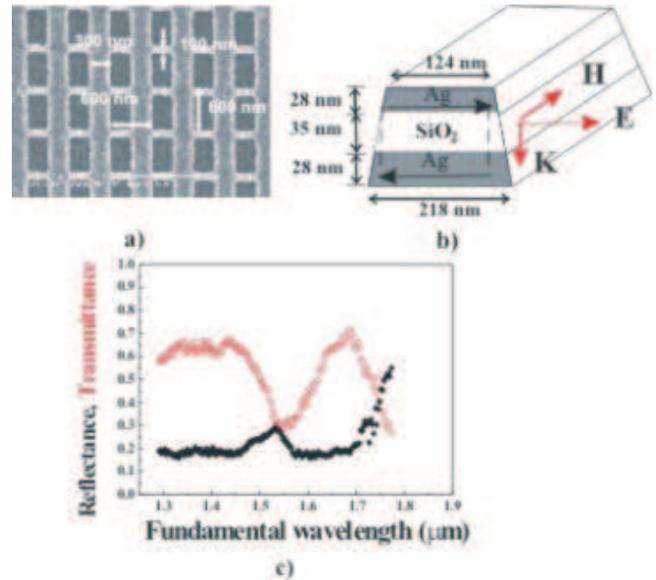} \caption{a) SEM image of the fishnet
structure; b) Schematic structure of the broad wire of "fishnet"
Ag/SiO$_{2}$/Ag structure. c) Linear transmittance (open dots) and
reflectance (solid dots) spectra. }\label{1}
\end{centering}
\end{figure}

The nonlinear optical spectroscopy of the fishnet structure was
performed using the tunable output from an optical parametric
system, pumped by the third-harmonic of a picosecond YAG:Nd laser.
The tunable IR beam was incident at 30$^{0}$ on the sample, and
the reflected  SHG, THG and four wave mixing signals were detected
by a photo-multiplier  and gated electronics system after spectral
filtering.

\begin{figure}[!h]
\begin{centering}
\includegraphics
[width=9.3cm]{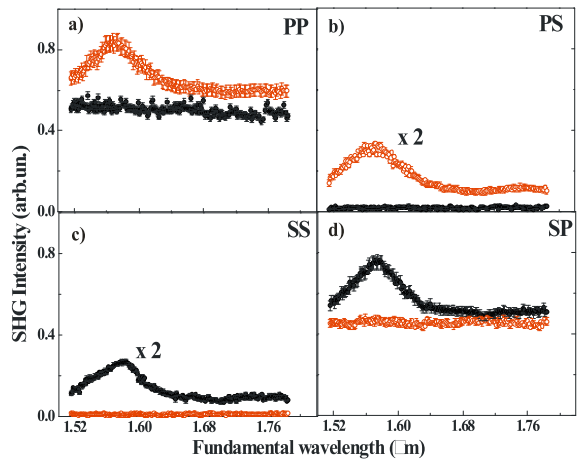} \caption{SHG spectra of the fishnet
structure for different polarization combinations: (a) p-in p-out,
(b) p-in s-out, (c) s-in s-out, (d) s-in p-out.  Solid dots and
open dots are for beam geometry with the incident magnetic field
component along and perpendicular to the broad Ag wire,
respectively. }\label{1}
\end{centering}
\end{figure}

The SHG spectra, obtained with the incident plane along (solid
dots) and perpendicular (open dots) to the broad Ag wires of the
fishnet are shown in Fig. 2. Polarization combinations employed
are indicated in the figure (PS, for example, denotes P and S
polarizations for the fundamental input and SH output,
respectively). To correct for the wavelength dependent incident
laser intensity, the SHG signals at each wavelength were
normalized to that of a smooth silver film with the PP
polarization combination. Two features are clear in the SHG
spectra. First, the SHG spectra display a resonance at 1.55 $\mu$m
whenever the input polarization has a magnetic-field component
along the  broad wires of the fishnet (i.e., P- or S-polarized
when the incident plane is perpendicular to or along the broad
wires, respectively) and the magnetic resonance is excited.
Otherwise, the SHG spectra are featureless with only non-resonant
contribution. Second, the PS and SS polarizations yield weaker SHG
signals with a much lower nonresonant background compared to PP
and SP. Presumably this is the result of symmetry. As in the case
of thin Ag films, SHG from a perfect fishnet structure, with the
incident plane coinciding with a mirror plane, is strictly
forbidden for PS and SS polarizations, but allowed for PP and SP.
That SHG with SS and PS is actually observable is an indication
that the fishnet sample is not ideally symmetric.

In Fig. 3, the THG spectra are shown for different sample
orientations and polarization combinations  using the same notations
as those for SHG in Fig. 2. Strong enhancement of THG is observed
when and only when the magnetic resonance is excited by the
fundamental beam. The stronger signal for PP and SS compared with PS
and SP polarization can also be understood from symmetry argument:
THG with PS and SP is forbidden in a perfect fishnet structure when
the incident plane coincides with a mirror plane. Compared with the
SHG spectra, the non-resonant contribution is smaller in the THG
signal and the resonance more pronounced.
\begin{figure}[!b]
\begin{centering}
\includegraphics
[width=9.3cm]{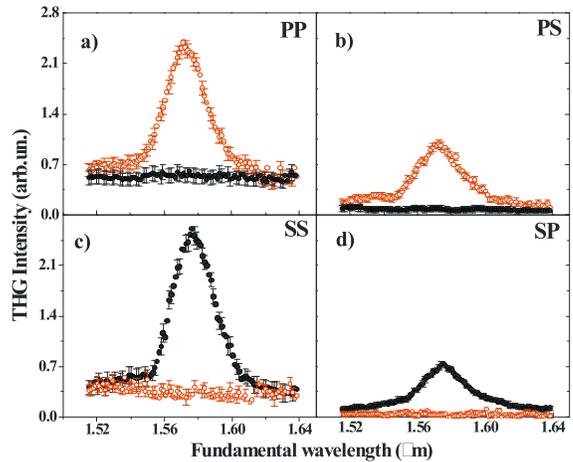} \caption{THG spectra of the fishnet
structure for different polarization combinations: (a) p-in p-out,
(b) p-in s-out, (c) s-in s-out, (d) s-in p-out. Solid dots and
open dots are for beam geometry with the incident magnetic field
component  along and perpendicular to the broad Ag wire,
respectively. }\label{1}
\end{centering}
\end{figure}

We compare in Fig. 4 the magnetic resonant features in the PP
spectra of linear absorption, SHG and THG. The resonance of THG is
clearly narrower than that of SHG, which is in turn sharper than
that of linear absorption. This is in striking contrast with
molecular resonances, where resonance lineshapes of harmonic
generation spectra are similar to that of linear absorption. It
indicates that resonances of "artificial molecules" in
metamaterials are characteristically different from those of
natural molecules. The difference arises because plasmon
resonances of metal nanostructures are intrinsically collective in
nature instead of local as in molecular transitions. Such
distinctions do not show up in linear optical properties, but
become obvious in nonlinear optical spectra where a resonant input
field participates multiple times in nonlinear processes. To
further establish the above-mentioned characteristic resonant
behavior in nonlinear responses of metamaterials, we measured
outputs of  four wave mixing at 3$\omega_{1}$,
2$\omega_{1}$+$\omega_{2}$ and $\omega_{1}$+2$\omega_{2}$ with
$\omega_{2}$ fixed (at wavelength of 1064 nm) and  $\omega_{1}$
scanned across the magnetic resonance. The observed spectra for PP
polarization are displayed in Fig. 4b with signals at
2$\omega_{1}$+$\omega_{2}$ and $\omega_{1}$+2$\omega_{2}$ scaled
up by 6.6 and 43, respectively.
\begin{figure}[!h]
\begin{centering}
\includegraphics
[width=8.5cm]{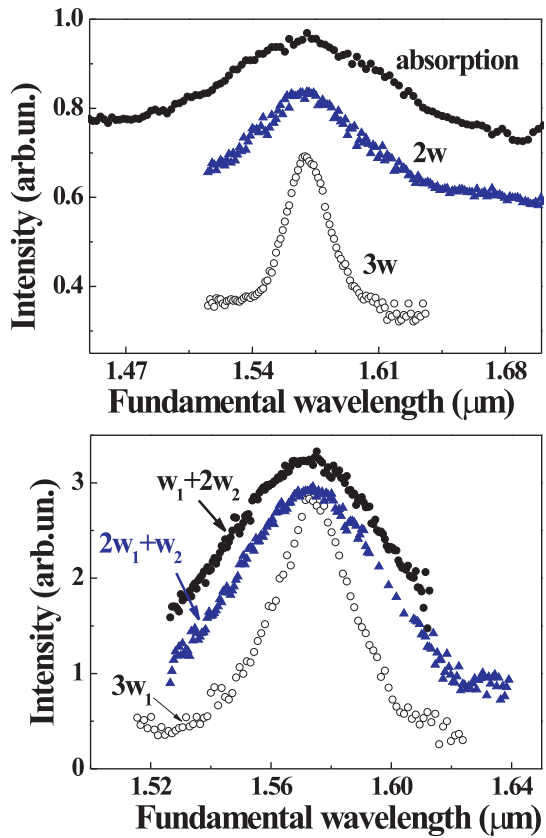} \caption{a) Comparison of SHG (triangle)
and THG (open dots) spectra in pp polarization combinations with
the linear absorption spectrum (solid dots). b) Spectra of
four-wave processes at 3$\omega_{1}$ (open dots),
2$\omega_{1}$+$\omega_{2}$ (triangles) and
$\omega_{1}$+2$\omega_{2}$ (solid dots). The signal strengths  at
2$\omega_{1}$+$\omega_{2}$ and $\omega_{1}$+2$\omega_{2}$ are
scaled up by 6.6 and 43, respectively, for ease of comparison.
}\label{1}
\end{centering}
\end{figure}
It is obvious that each additional $\omega_{1}$ component in the
input leads to extra enhancement of the output at resonance, and
the corresponding resonance peak becomes increasingly sharper.

For better understanding of the observed results, we realize that
the real source of \emph{n}th harmonic generation in metamaterials
is the induced effective electric dipole on each nanostructural
unit:

\begin{equation}
\label{eq1} \vec{P}^{(n)}=\int_{V}\tilde{L}(\vec{r},
n\omega):\tilde{\chi}^{(n)}(\vec{r}):[\vec{E}_{loc}(\vec{r},\omega)]^{n}dV,
\end{equation}
where the integration is over the volume of the unit,
$\tilde{\chi}^{(n)}$ is the local \emph{n}th-order nonlinear
susceptibility and
$\vec{E}_{loc}(\vec{r},\omega)=\tilde{L}(\vec{r},\omega):\vec{E_{0}}(\omega)$
is the local field with $\tilde{L}(\vec{r},\omega)$ being the
local field correction factor and $\vec{E_{0}}(\omega)$ the input
field. For simplicity, we have neglected the contribution of
magnetic-dipole and electric-quadrupole to $\vec{P}^{(n)}$ in
Eq.(1). The harmonic output with polarization along $\hat{n}$ is
given by $S^{(n)}\propto|\hat{n}\cdot\vec{P}^{(n)}|^{2}$. The
local field here can be decomposed into two components, one
associated with the magnetic resonance and the other not. The
resonant component is relatively enhanced when approaches
resonance. Thus we can write the local field correction factor as
$\tilde{L}(\vec{r},\omega)=\tilde{A}(\vec{r},\omega)+\tilde{B}(\vec{r},\omega)/D(\omega)$
with $D$ approximated by $D=\omega-\omega_{0}+i\Gamma$ describing
the magnetic plasmon resonance.

The symmetry requirement for harmonic generation is naturally
incorporated in the volume integration of Eq.(1), which, for
example, vanishes for symmetry forbidden processes in a perfect
fishnet structure. The integral depends sensitively on the field
distribution, and nonresonant and resonant terms respond
differently to nanostructure change because of their different
local field distributions: For a symmetry-forbidden harmonic
generation process, if symmetry-breaking modification of the
nanostructure is not severe, the nonresonant signal is still
expected to be small. At resonance, however,
$\tilde{B}(\vec{r},\omega)$ being different from
$\tilde{A}(\vec{r},\omega)$ can have a spatial distribution
emphasizing contribution from the symmetry-breaking part of the
structure ~\cite{ss}, thus generating a relatively strong resonant
harmonic output. This explains our observation of resonant spectra
with very weak nonresonant background for symmetry-forbidden
harmonic generation processes in the fishnet structure presented
in Figs. 2 and 3. Numerical calculation on a realistic fishnet
structure hopefully will quantify the different effects of
symmetry breaking on resonant and nonresonant harmonic generation.

If the nonresonant part of $\tilde{L}(\vec{r},\omega)$ could be
neglected, we would have, for the nth harmonic generation,
$\vec{P}^{(n)}\propto|D|^{-n}$ and $S^{(n)}\propto|D|^{-2n}$. With
the presence of the nonresonant $\tilde{A}(\vec{r},\omega)$ term
in $\tilde{L}(\vec{r},\omega)$, the signal $S^{(n)}$ now has terms
of $|D|^{-2n}$ with $m = 0,1,\cdot\cdot\cdot,n$, and its resonant
lineshape often appears broader than $|D|^{-2n}$. However, the
$|D|^{-2n}$ term is always significant, making the THG ($n = 3$)
spectrum sharper than that of SHG ($n = 2$); similar behavior is
seen in comparing spectra of wave mixing at 3$\omega_{1}$,
2$\omega_{1}$+$\omega_{2}$ and $\omega_{1}$+2$\omega_{2}$.

In summary, we have measured spectra of SHG, THG and four-wave
mixing from a fishnet metamaterial around its magnetic resonance
with different input/output polarization combinations. The results
show that the resonant enhancement is much stronger in nonlinear
optical response than that in the linear case, and the more times
the resonant input field participates in the mixing process, the
sharper the resonant spectrum appears to be.  This is because the
resonance is plasmonic in nature and shows up in the local-field
correction factors in the nonlinear responses. Such resonant
behavior is expected to appear in all nonlinear optical processes
involving plasmon resonances in metamaterials.

\begin{acknowledgments}
This work was initiated under the sponsorship of DARPA. The
authors were supported by Director, Office of Science, Office of
Basic Energy Sciences, Materials Sciences and Engineering
Division, of the U.S. Department of Energy under Contract No.
DE-AC03-76SF00098. FW acknowledges a fellowship support from the
Miller Institute of the University of California.

\end{acknowledgments}


\begin{thebibliography}{26}
\bibitem {ss} For the SS case with incident plane along the broad Ag wire,
for example, it can be seen that the resonant field distribution
emphasizes the field at the edges of the wires where the symmetry
breaking occur, but the nonresonant field distribution does not.
\end{thebibliography}
\end{document}